\renewcommand{\vec}[1]{\boldsymbol{#1}}
\documentclass[prl,showpacs,floatfix,twocolumn]{revtex4}

\usepackage{graphicx}
\usepackage{amsmath}

\begin{document}

\title{Qubit state monitoring by
 measurement of three complementary observables}

\author{Rusko Ruskov,$^{1,}$\footnote{On leave from INRNE, Sofia BG-1784, Bulgaria.}
%Institute for Nuclear Research and Nuclear Energy, Sofia BG-1784, Bulgaria}
Alexander N. Korotkov,$^2$ and Klaus M{\o}lmer$^1$}
\affiliation{
$^1$Lundbeck Foundation Theoretical Center for Quantum System Research,
Department of Physics and Astronomy,
Aarhus University, DK-8000 Aarhus C, Denmark \\
$^2$Department of Electrical Engineering, University of California,
Riverside, CA 92521, U.S.A. }

\date{\today}

\begin{abstract}
We consider the evolution of a spin $1/2$ (qubit) under the
simultaneous continuous measurement of three non-commuting qubit operators
$\hat{\sigma}_x$, $\hat{\sigma}_y$, $\hat{\sigma}_z$. For identical ideal detectors the
qubit state evolves by approaching a pure state with a random
direction in the Bloch vector space and by undergoing locally isotropic
diffusion in the perpendicular directions. The quantum state conditioned on the
complete detector record is used to assess the fidelity of
classically inspired estimates based on running time averages and
discrete time bin detector outputs.

\end{abstract}

\pacs{03.65.Ta, 03.67.-a, 42.50.Dv, 02.50.Tt}
\maketitle

%{\it Introduction.}
The needs of quantum computing/communication \cite{NielsenChuang}  %,WisemanMilburn-control}
are stimulating rapid progress in control of single quantum systems.
Recent experiments demonstrate coherent manipulation of quantum
systems, including Rabi oscillations and entangling operations with
few qubits.
   An important direction for advanced quantum control is
to realize continuous monitoring of a quantum system.
    Theory of
continuous quantum measurement
\cite{Belavkin,Carmichael,WisemanMilburn93,DalibardCastinMolmer,Kor-99-01,RusKorMiz-BellLegg}
and experiments \cite{Katz-Sci-prl,Haroche-exp,Rabi-peak,Orozco} have been
carried out on a number of systems.
   The quantum monitoring can be
used to prepare highly pure states and
entangled states
\cite{WisemanMilburn93,RusKor-1qb2,   %RusKor-fb,RusKor-ent,
%Kor-fb-simple,
KJacobs,WisemanRalph,Negretti,Combes} and for continuous error
correction \cite{Ahn}.

A particularly interesting case is when non-commuting variables are
being measured simultaneously.
In Ref.\ \cite{JordanButtiker}  %\onlinecite{JordanButtiker}
the signal cross-correlation for two such detectors of an evolving
qubit was calculated.
  In Ref.\ \cite{WeiNazarov}        %\onlinecite{WeiNazarov}
Wei and Nazarov considered measurement outcomes for three detectors
measuring a qubit in orthogonal directions.
They analyzed the statistics
of the integrated outcomes $v_k$ for each detector and showed that
if these outcomes happen to be sufficiently large, then the
normalized vector $\vec{v}/|\vec{v}|$
is close to the Bloch vector of the actual qubit state.

\begin{figure}
\vspace*{-0.3cm} \centering
\includegraphics[width=1.7in]{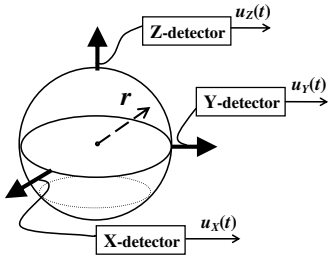}
\vspace*{-0.2cm}
\caption{A qubit measured by three orthogonal detectors.}
%{\bf (Change I to u?)}
\label{schematic}
\end{figure}

In this Letter we consider simultaneous continuous measurement of
the qubit observables $\hat{\sigma}_x, \hat{\sigma}_y, \hat{\sigma}_z$,
illustrated in Fig.\ 1.
The setup can be in principle realized with a trapped atom
probed dispersively by optical cavity fields. In contrast to \cite{WeiNazarov},
we explicitly take into account the qubit evolution due to measurement,
and analyze the problem of monitoring the qubit state using the
measurement records. The three Pauli observables are complementary,
but the incremental changes of the quantum state due to the weak
measurements carried out in infinitesimal time intervals commute,
and the simultaneous measurements contribute to purification of the quantum state
three times faster than if only a single observable is measured.
While the measurements drive the system towards the Bloch sphere
surface (pure states),
they cause locally isotropic diffusion in angular directions.

If the observer has access only to detector read out signals integrated
over finite time intervals, the back actions associated with these accumulated signals
do not commute, and the state of the qubit can only be approximately determined.
The quality of the state estimate in this case is thus a measure of the role of
complementarity of the observables detected. Comparing the exact qubit evolution
with simple classically inspired ways
of monitoring, we show that running averages with an exponential window
can provide the fidelity of state monitoring up to 0.94. We also
show that if the available measurement record is averaged over
discrete time steps $\Delta t$, the monitoring fidelity decreases
with $\Delta t$ quite slowly.

{\it Model.}
   We consider continuous measurement of the qubit observables
$\hat{\sigma}_x$, $\hat{\sigma}_y$, and $\hat{\sigma}_z$  by three
linear detectors with output signals $I_k(t)$.   %, k=x,y,z$.
Let $u_k(t)=I_k(t)-I_{0,k}$,
where $I_{0,k}$ denotes the uniform average of the outcomes of detector $k$ over the
two qubit states, and let $\Delta u_k$ denote the detector responses, i.e.,
the difference of the mean signal for the qubit states. We can then write
\begin{equation}
u_k(t)  = \frac{\Delta u_k}{2} \,
\mbox{Tr}[\hat{\rho}(t)\,\hat{\sigma}_k] + \xi_k(t), \,\,\, k=x,y,z.
\label{det-currents}
\end{equation}
The $\mbox{Tr}[\hat{\rho}\hat{\sigma}_k]$ expectation values, determined by the
time dependent qubit density matrix $\hat{\rho}$, are in the following represented as the Bloch
vector $\vec{r}=(x,y,z)$. In Eq.(\ref{det-currents}) $\xi_k(t)$ are independent
white noises with one-sided spectral densities $S_k$,
$\langle\xi_k\xi_l dt\rangle = \delta_{kl} S_k/2$.
    The qubit evolution due to measurement by a linear detector
(amplifier with infinite gain) can be described by three parameters \cite{Kor-99-01,Clerk}:
the so-called measurement time $\tau_{{\rm meas},k}=2S_k/(\Delta u_k)^2$,
which determines the rate of quantum (informational)
back-action, a factor $K_k$ describing the classical back-action
correlated with the output noise $\xi_k$, and an ensemble dephasing
rate $\Gamma_k$, related to the single-qubit dephasing rate
$\gamma_k$ as $\Gamma_k=\gamma_k+ 1/2\tau_{{\rm meas},k}+K_k^2S_k/4$.
In this paper we are interested in the quantum    %mechanical
back action due to measurements,
and we assume
the absence of classical back-action, $K_k=0$ as well as the absence of any Hamitonian
driving of the qubit.

    If the measurement is performed only by one
$\hat{\sigma}_k$-detector ($\Gamma_l=0$ for $l\neq k$), then the
probability density of its integrated result
$\bar{u}_k(\tau)=\tau^{-1} \int_0^{\tau} u_k(t) dt$ is
$P_{\rm tot}(\bar{u}_k)=\sum_i \rho_{ii}(0) P_i(\bar{u}_k)$, where the qubit
density matrix $\hat\rho$ is written in the $\hat{\sigma}_k$ basis
($i=1,2$) and
$P_i(\bar{u}_k)=\sqrt{\tau/\pi S_k} \exp\{-[\bar{u}_k+(-1)^i \Delta u_k/2 ]^2 \tau/S_k\}$
are the Gaussian
distributions for the basis states.
Then the qubit evolution is given by
the Bayesian quantum filter \cite{Kor-99-01}
 \begin{equation}
\rho_{ij}(\tau) = \rho_{ij}(0) e^{-\gamma_k\tau (1-\delta_{ij})}
\sqrt{P_i(\bar{u}_k)P_j(\bar{u}_k)}\, /\, P_{\rm tot}(\bar{u}_k) .
\label{quantBayes-z}
 \end{equation}

{\it Qubit evolution with three detectors.} In the case of three
detectors measuring the qubit in the orthogonal bases, it is
impossible to use the quantum Bayes rule for a finite $\tau$ because
the measurement back-actions do not commute with each other.
Therefore we should apply Eq.~(\ref{quantBayes-z}) in the
differential form (for small $dt$) in the three orthogonal bases
corresponding to the measured observables and then sum up the
contributions to the qubit evolution. In this way we obtain the
following equation in the Stratonovich form for the $x$-component of
the qubit Bloch vector $\vec{r}(t)$ given the measurement record
$u_k(t)$:
\begin{eqnarray}
&& \dot{x} = (1-x^2) (\Delta u_x/S_x) u_x - x y (\Delta u_y /S_y) u_y
      \nonumber\\
&& \qquad - x z (\Delta u_z/S_z )  u_z - (\gamma_y + \gamma_z)  x .
   \label{3det-Stratx}
\end{eqnarray}
Evolution equations for the components $y$ and $z$ can be obtained
by cyclic permutation of variables in Eq.~(\ref{3det-Stratx}).

{\it Identical detectors.} In what follows we consider the case of
three identical detectors: $\Delta u_k/S_k = \Delta u/S = a$ (we
assume $a>0$) and $\gamma_k=\gamma\geq 0$. Then the qubit evolution
(\ref{3det-Stratx}) can be rewritten in a vector form as
  \begin{equation}
\dot{\vec{r}} = -2 \gamma\,\vec{r}
+ a \left\{\vec{u}\,(1-r^2)
- \left[\vec{r} \times
\left[\vec{r}\times \vec{u}\right]\right] \right\}
\label{Strat-vec2} ,
  \end{equation}
where  $\vec{u}\equiv (\Delta u/2)\vec{r}+\vec{\xi}(t)$ is the
vector of results, Eq.~(\ref{det-currents}), and $r=|\vec{r}|$.
The evolution (\ref{Strat-vec2}) is invariant under arbitrary
rotations and can be represented as evolution due to one-detector
measurement along fluctuating random direction of $\vec{u}$. It is
interesting to note that while measurement of only single observable
$\hat{\sigma}_k$ ``attracts'' the qubit state to one of the
corresponding eigenvectors, the simultaneous measurement of
$\hat{\sigma}_x, \hat{\sigma}_y, \hat{\sigma}_z$ leads to no
preferable direction in the Bloch space.

    The ensemble-averaged evolution is also isotropic:
$\dot{\vec{r}}=-2\Gamma\, \vec{r}$, that is easier to see from the
It\^{o} form \cite{Gardiner} of (\ref{Strat-vec2}):
  \begin{equation}
\dot{\vec{r}} = -2 \Gamma\,\vec{r} + a \left\{\vec{\xi}\,(1-r^2) -
\left[\vec{r}\times \left[\vec{r}\times \vec{\xi}\right]\right]
\right\} , \label{Ito-vec2}
  \end{equation}
where $\Gamma=\gamma+ \Gamma_0$ is the one-detector  ensemble decoherence and
$\Gamma_0=(\Delta u)^2/4S=1/2\tau_{\rm meas}$. We also introduce the
efficiency (ideality) of the measurement $\eta = \Gamma_0/\Gamma$.

Transforming Eq.\ (\ref{Ito-vec2}) to polar coordinates, we obtain
the following evolution for the radial component $r$:
\begin{equation}
\dot{r} = 2\Gamma_0\, (1/r-r/\eta ) + a(1-r^2) \, \xi_r
\label{Ito-radial} ,
\end{equation}
where $\xi_r(t)=\vec{e_r} \cdot \vec{\xi}(t)$ is the noise component
along $\vec{r}$ with the same spectral density:
$\langle\xi_r \xi_r\, dt\rangle = S/2$.
   In  directions perpendicular to $\vec{r}$ Eq.\ (\ref{Ito-vec2})
leads to a locally isotropic Brownian diffusion with coefficient $a$;
%{\bf (notice that $a$ decreases with increasing the output noise $S$)};
correspondingly, the angular evolution of the projection onto
the Bloch sphere surface has the diffusion coefficient $a/r$ (this
can be shown by using locally geodesic coordinates).

    In particular, for ideal measurement ($\eta=1$) and pure initial
state, the state remains pure [$r=1$, see Eq.\ (\ref{Ito-radial})]
and the diffusion
on the Bloch sphere can be described by the
Fokker-Planck (FP) equation $\partial p(\theta,\varphi)/\partial
t=\Gamma_0 \Delta_{\theta,\varphi}p(\theta,\varphi)$ where
$\Delta_{\theta,\varphi}$ is the angular part of the Laplacian. The
solution of this equation \cite{Perrin} at time $\tau$ is
$p(\Theta;\tau)=\sum_{n=0}^{\infty}\frac{2n+1}{4\pi}e^{-n(n+1)V/4}P_n(\cos\Theta)$,
where $\Theta$ is the angle from the
initial state, $P_n(z)$ are the Legendre polynomials, and
$V=4\Gamma_0 \tau=2\tau/\tau_{\rm meas}$ is the variance. Obviously,
for $\tau \gg \tau_{\rm meas}$ the initial state is  forgotten, and
the distribution $p(\Theta;\tau) \rightarrow 1/4\pi$ becomes
isotropic. We note that while the average state approaches the
center of the Bloch sphere, the actual monitored qubit state remains
pure, performing a random walk on the sphere.
We also note that the back action quantified by the diffusion $\Gamma_0$ decreases with
increasing output noise $S$.

{\it Purification dynamics.}
   As seen from Eq.\ (\ref{Ito-radial}), if $r< \eta^{1/2}$, then on
average $\dot{r}>0$, which means state purification. Uncertainty of
our knowledge about the state is characterized by the linear entropy
$S_{lin}=1-{\cal P}$, where
${\cal P}\equiv 2 \mbox{Tr}\hat{\rho}^2 - 1 = r^2$ is the state purity.
   For an ideal measurement, $\eta=1$,
and starting from a non-pure initial state, the qubit will purify
(${\cal P}\rightarrow 1$) on a time scale of the order of $\tau_{\rm meas}$.
  For a non-ideal measurement, $\eta < 1$, purity will
continue to fluctuate around a stationary average value,
$\langle{\cal P}\rangle_{st} < 1$.

To analyze the purification dynamics we use Eq.~(\ref{Ito-radial})
to derive It\^{o} equation for the purity:
 $d{\cal P}/dt = 2\Gamma_0 [2 (1-{\cal P}/\eta ) + (1-{\cal P})^2]+ 2 a (1-{\cal P})
\sqrt{{\cal P}} \xi_r(t)$.
The corresponding FP equation\cite{Gardiner} is
$\frac{\partial p({\cal P},t)}{\partial t} =
-\frac{\partial}{\partial {\cal P}} \left[ A({\cal P}) p({\cal P},t)\right]
+ \frac{1}{2}\,\frac{\partial^2}{\partial {\cal P}^2}\,
\left[ B({\cal P}) p({\cal P},t)\right]$
with coefficients
$A({\cal P}) = 2 \Gamma_0 [2 (1-{\cal P}/\eta ) + (1-{\cal P})^2]$,
$B({\cal P}) = 8 \Gamma_0 {\cal P} (1-{\cal P})^2$, and initial distribution
$p({\cal P},0) = \delta({\cal P}-{\cal P}_0)$.
   At $t \gg \tau_{\rm meas}$ the purity  reaches a
stationary distribution
   \begin{equation}
p_{st}({\cal P},\eta) = N^{-1}\, \frac{\sqrt{{\cal P}}}{(1-{\cal P})^3}\,
\exp{\left[-\frac{{\cal P} (1-\eta)}{(1-{\cal P})\eta}\right]} ,
  \label{FP-Pur-stationary}
\end{equation}
where $N$ is the normalization.
    For $\eta\rightarrow 1$, $p_{st}({\cal P},\eta)$ approaches the
$\delta$-function at ${\cal P}=1$.

In Fig.\ 2 we show the average purity (solid lines)
$\langle {\cal P}\rangle_{FP} (t)=\int_0^1 {\cal P}\, p({\cal P},t) \, d{\cal P}\,$
for measurement efficiencies $\eta=1$, 0.5, and 0.1, calculated by
numerically solving the FP equation starting from the Bloch sphere center
(${\cal P}_0=0$). The FP distribution $p({\cal P},t)$ (not shown)
has been also confirmed by the simulations using
Eq.~(\ref{Strat-vec2}).

\begin{figure}
%\vspace*{-0.4cm}
\centering
\includegraphics[width=2.8in]{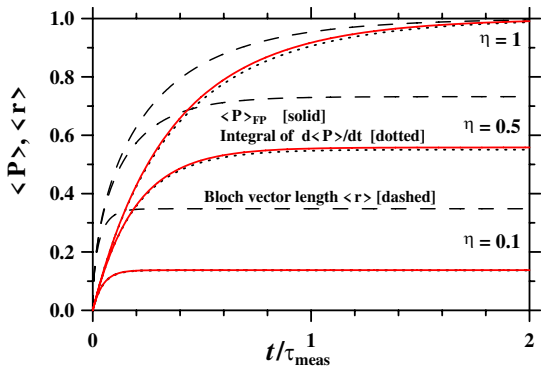}
\vspace{-0.2cm}
\caption{Solid lines: averaged purity evolution $\langle {\cal
P}\rangle_{FP}(t)$  starting from the fully mixed state. Dotted
lines: same evolution integrating Eq.~(\ref{av-pur-rate}). Dashed
lines: evolution of $\langle r\rangle = \langle\sqrt{\cal
P}\rangle_{FP}$.
} \label{purity_evolution}
\end{figure}

The purification dynamics can be approximated using the
ensemble-averaged purification rate \cite{KJacobs,WisemanRalph}
obtained from the above It\^{o} equation, starting from a given
purity:
    \begin{equation}
\langle d{\cal P}\rangle/dt = 2 \Gamma_0 [2 \left(1-{\cal P}/\eta \right)
+ (1-{\cal P})^2 ] \label{av-pur-rate}  .
\end{equation}
For ideal detectors ($\eta=1$) this becomes $2 \Gamma_0 (1-{\cal P})\, (3-{\cal P})$,
that can also be easily obtained from the
purification result \cite{KJacobs,JordanKor}
$\langle d{\cal P}\rangle/dt = 2 \Gamma_0 (1-{\cal P})\, (1-z^2)$ for a $z$-detector
by adding the contributions from measurements in $x$ and $y$
directions, so that $1-z^2\rightarrow 3-r^2$. The purification by 3
detectors probing in mutually unbiased bases is on average 3 times
faster than for a $z$-detector (averaging is over the directions in
the Bloch space), since $ \overline{z^2} =r^2/3$. Also, the
3-detector purification rate $\langle d{\cal P}\rangle/dt$ is
isotropic, in contrast to the single-detector case, for which the
adaptive measurement perpendicular to the spin vector leads to the
fastest purification \cite{KJacobs,WisemanRalph} (Ref.\ \cite{Combes}
demonstrates that non-adaptive switching between
random bases performs comparatively well).
   It is important to note that the average purity
$\langle {\cal P}\rangle_{FP}(t)$ differs from the naive  integration of
Eq.~(\ref{av-pur-rate}) (dotted lines in Fig.\ 2) because
$\langle d{\cal P}\rangle \neq d\langle{\cal P}\rangle$ and the purity
distribution $p({\cal P},t)$ is generally different from
$\delta$-function. In particular,  $\langle{\cal P}\rangle_{st}$ is
slightly higher (for $\eta\neq 0,1$) than the stationary value
$(1 + 1/\eta) - \sqrt{(1 + 1/\eta)^2 - 3}$ derived from
Eq.\ (\ref{av-pur-rate}).

{\it Classically inspired state monitoring.} Exact monitoring of the
qubit state is realized by integrating the evolution
Eq.~(\ref{Strat-vec2}) given the measurement record $u_k(t)$.
 However, such real-time computation may be a challenge
experimentally, and therefore it is interesting to analyze the fidelity
of simplified signal processing algorithms.
   To decrease the noise component and reduce the bandwidth of
signals given by Eq.\ (\ref{det-currents}), it is natural to average
them over a running time-window:
$\tilde{u}_k(t) \equiv \int_{-\infty}^t g(t-t') u_k(t')\, dt' $,
where $g(\Delta t)$ is the window profile. We have considered (i) a
rectangular window of duration $\tau$: $g(\Delta t)=\tau^{-1}$ for
$\Delta t < \tau$ and zero otherwise, and (ii) an exponential window
with decay time $\tau$:
$g(\Delta t)=\tau^{-1}\exp (-\Delta t/\tau)$.
  The analyzed monitoring algorithm is very simple: at any $t$
in the stationary regime ($t \gg \tau,\tau_{\rm meas}$) we estimate
the qubit state as the pure state
$\vec{r}_{\rm est}(t)=\vec{\tilde{u}}(t)/|\vec{\tilde{u}}(t)|$.
  The algorithm fidelity is defined as the time-averaged scalar
product of this vector with the actual state $\vec{r}(t)$:
\begin{equation}
F \equiv 2\,\langle \mbox{Tr} \hat{\rho}_{\rm est}
\hat{\rho}\rangle_t - 1 = \langle \vec{r} \cdot
\vec{\tilde{u}}/|\vec{\tilde{u}}| \rangle_t .
 \label{long-t-fidelity}
\end{equation}

In Fig.\ 3 we show the fidelity $F$ vs.\ the window  duration $\tau$
for the rectangular (solid lines) and exponential (dashed lines)
windows, calculated by simulating the evolution (\ref{Strat-vec2})
for $\eta=1$, 0.5, and 0.1.
 For $\eta=1$ the fidelity reaches a maximum of $F_{\rm max}=0.94$ for the exponential
window with $\tau=0.6\,\tau_{\rm meas}$ (for the rectangular window
$F_{\rm max}=0.87$ at $\tau=0.9\,\tau_{\rm meas}$).
    For small $\tau$ the fidelity is suppressed
due to large contribution from fluctuations:
$|\vec{\tilde{u}}| \sim \tau^{-1/2}$,
while for $\tau\agt \eta \tau_{\rm meas}$ it
decreases because signals from distant past loose their relevance to
$\vec{r}(t)$. To analyze the latter effect quantitatively, we have
used Eq.\ (\ref{Strat-vec2}) to find the signal-qubit correlations:
$\langle u_z(t-\Delta t) z(t)\rangle=(\Delta u/2) \exp (- \Delta t/\eta\tau_{\rm meas})$,
$\langle u_z(t-\Delta t ) x(t)\rangle=\langle u_z(t-\Delta t) y(t)\rangle =0$; other
correlators are similar \cite{isotropy}.

\begin{figure}
%\vspace*{-0.4cm}
\centering
\includegraphics[width=2.8in]{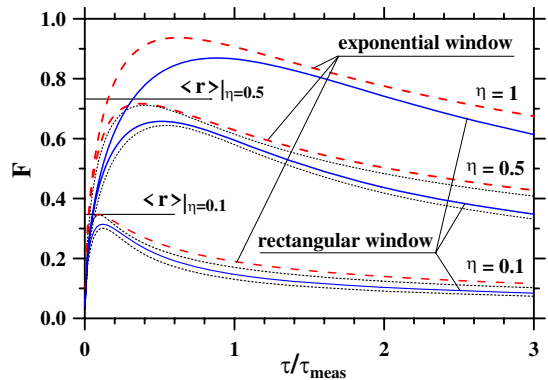}
\vspace{-0.2cm}
\caption{%Long-time averaged
Monitoring fidelity (\ref{long-t-fidelity}) vs.\ window duration
$\tau$ for the rectangular (solid lines) and exponential windows
(dashed lines). Dotted lines show the uncorrelated product
$\langle r\rangle \langle \cos \phi\rangle_t$. }
\label{fidelity_simple_alg}
\end{figure}

Since $F =\langle r \cos\phi \rangle_t$, where $\phi$ is the angle
between $\vec{\tilde u}$ and $\vec{r}$, the fidelity is bounded from
above by the stationary Bloch vector length $\langle r\rangle$
reached at $t\rightarrow \infty$ (Fig.\ 2, dashed lines).
  In Fig.~3 these bounds are shown as horizontal lines:
$\langle r\rangle=0.732$ for $\eta =0.5$ and $\langle r\rangle=0.348$
for $\eta =0.1$ (obviously, $\langle r\rangle=1$ for $\eta =1$).
 It is interesting to see that with decreasing $\eta$,
the exponential-window $F_{max}$ approaches $\langle r\rangle$.
  This means that at optimal $\tau$ either $\vec{\tilde u}$
becomes practically aligned with $\vec{r}$, $\langle
\cos\phi\rangle_t \to 1$, or there is a significant correlation
between fluctuations of $r(t)$ and $\phi (t)$. This correlation can
be checked by comparing $F$ with the uncorrelated value $\langle
r\rangle \langle \cos\phi\rangle_t$ shown in Fig.\ 3 by dotted
lines. We see practically no correlation at the optimal point for
small $\eta$, which means that $\langle \cos\phi\rangle_t \to 1$ (we
have also checked this fact directly).

\begin{figure}
%\vspace*{-0.4cm}
\centering
\includegraphics[width=2.8in]{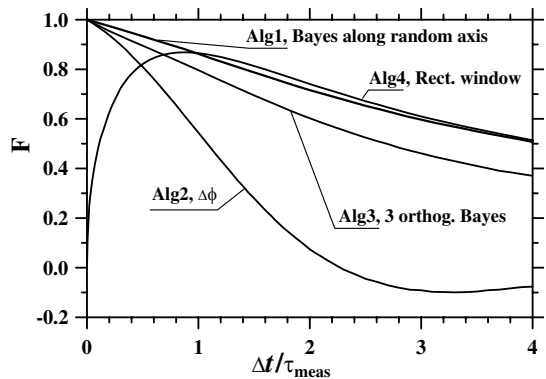}
\vspace{-0.2cm}
\caption{
Monitoring fidelity vs.\ time step $\Delta t$ for four algorithms
processing time-discretized detector data (see text). }
\label{fidelity_algorithms}
\end{figure}

{\it Algorithms for time-discretized data.} We consider now a
situation when only the integrated detector signals
$u_{k}^{(n)} \equiv \frac{1}{\Delta t} \int_{t_n-\Delta t}^{t_n} u_k(t')\, dt'$
are available at discrete time moments $t_n = n \Delta t$,
$n=0,1,\ldots$, as in the experiment \cite{Rabi-peak}. Then exact
monitoring of the qubit state is obviously impossible, especially if
$\Delta t \sim \tau_{\rm meas}$, and it is interesting to analyze
performance of various monitoring algorithms in this case. Assuming
ideal detectors to deal with pure states only, we have studied four
algorithms, which output estimated states $\vec{r}_{\rm est}$ at
moments $t_n$, and calculated their fidelities (shown in Fig.\ 4)
defined as $F=\langle\vec{r}\cdot \vec{r}_{est}\rangle_n$, similar
to Eq.\ (\ref{long-t-fidelity}) but with averaging over moments
$t_n$. The actual evolution $\vec{r}(t)$ in this case is simulated
via Eq.\ (\ref{Strat-vec2}), and we have checked that the fidelity
does not depend on the inaccuracy of the initial state estimate.
  {\it Algorithm 1} treats the vector of measurement data
$\vec{u}^{(n)}$ as a single measurement of the spin component along
this vector and updates the qubit state using the quantum Bayes rule
(\ref{quantBayes-z}) in the corresponding basis, changing at each
time step. For small $\Delta t$ the fidelity of this algorithm is
$F \approx 1-0.14 \Delta t/\tau_{\rm meas}$. Somewhat unexpectedly,
even for $\Delta t \simeq \tau_{\rm meas}$ the fidelity is still
quite good.
   {\it Algorithm 2} at each step rotates the previous Bloch vector
$\vec{r}(t_{n-1})$ towards the vector $\vec{u}^{(n)}$ by the angle
$\Delta\phi=(\Delta u/S) u^{(n)}_\perp\ \Delta t$, which is
determined by the component $u^{(n)}_\perp$ of the vector
$\vec{u}^{(n)}$ perpendicular to $\vec{r}(t_{n-1})$. Even though for
small $\Delta t$ this is very similar to the Algorithm 1, the
fidelity decreases more rapidly with increasing $\Delta t$.
   {\it Algorithm 3} treats
the three measurement outcomes as the results of sequential
measurements of the three spin components and uses the Bayesian
update rule accordingly. For $\Delta t \to 0$ the fidelities of all
three algorithms approach unity, though with different slopes.
   {\it Algorithm 4} treats the data available at moments $t_n$ in
the same way as the running rectangular window (see the upper solid
line in Fig.~3) and estimates the state as
$\vec{u}^{(n)}/|\vec{u}^{(n)}|$.
Algorithm 4 suffers from large
statistical errors for short $\Delta t$; however, for
$\Delta t \approx \tau_{\rm meas}$ the fidelities of algorithms 1 and 4 become
practically equal, and for longer $\Delta t$ the Algorithm 4 becomes
the best among considered algorithms.

In conclusion, state monitoring and purification by simultaneous
measurements of non-commuting observables has been described by
quantum filtering theory. The shortcomings of simple, effective,
algorithms reflect the difficulty of estimating quantum states
from incomplete measurement data.
The incompleteness of the time averaged or integrated data is due to
complementarity and the non-commuting back action operations in the coarse
grained limit
of finite sampling times. Our analysis shows this very clearly and it
quantifies the approach to perfect state estimation in the limit of
continuous measurement and quantum filtering.

The authors thank Yuli V. Nazarov and Hongduo Wei for useful
discussions.
A.N.K. was supported by NSA/IARPA/ARO
grant W911NF-08-1-0336.


\begin{thebibliography}{99}
\bibitem{isotropy}

The cross-correlators vanish since  %because
%$\langle xy\rangle\! =\! \langle xz\rangle\! =\! \langle yz\rangle=0$
$\langle xy\rangle=\langle xz\rangle=\langle yz\rangle=0$
from symmetry, while
correlators like $\langle u_z(t-\Delta t) z(t)\rangle$ have two contributions:
$\langle z(t-\Delta t) z(t)\rangle = \frac{1}{3}\exp (-2\Gamma \Delta t)$ and
$\langle \xi_z(t-\Delta t) z(t)\rangle = \frac{\Delta u}{3}\exp (-2\Gamma \Delta t)$.
\end{thebibliography}
\end{document}